# Revealing superconducting gap in $La_3Ni_2O_{7-\delta}$ by Andreev reflection spectroscopy under high pressure

Jianning Guo[1,†], Yuzhi Chen[1,†], Yulong Wang[1], Hualei Sun[2], Deyuan Hu[3], Meng Wang[3,*], Xiaoli Huang[1,*] and Tian Cui[1,4]

[1] *State Key Laboratory of High Pressure and Superhard Materials, College of Physics, Jilin University, Changchun 130012, China*

[2] *School of Sciences, Sun Yat-sen University, Shenzhen 518107, China*

[3] *Center for Neutron Science and Technology, Guangdong Provincial Key Laboratory of Magnetoelectric Physics and Devices, School of Physics, Sun Yat-Sen University, Guangzhou, China.*

[4] *School of Physical Science and Technology, Ningbo University, Ningbo 315211, China*

†These authors contributed equally to this work
*Corresponding authors, emails:
huangxiaoli@jlu.edu.cn (X. Huang)
wangmeng5@mail.sysu.edu.cn (M. Wang)

**Abstract:**

The recent discovery of compressed superconductivity at 80 K in $La_3Ni_2O_{7-\delta}$ has brought nickelates into the family of unconventional high-temperature superconductors. However, due to the challenges of directly probing the superconducting pairing mechanism under high-pressure, the pairing symmetry and gap structures of nickelate superconductors remain under intense debate. In this work, we successfully determine the microscopic information on the superconducting gap structure of $La_3Ni_2O_{7-\delta}$ samples subjected to pressures exceeding 20 GPa, by constructing different conductance junctions within diamond anvil cells. By analyzing the temperature-dependent differential conductance spectra within the Blonder-Tinkham-Klapwijk (BTK) model, we have determined the superconducting energy gap at high pressure. The differential conductance curves reveal a two-gap structure with $\Delta_1 = 23$ meV and $\Delta_2 = 6$ meV, while the BTK fitting consistent with an *s*-like, two-gap spectrum. The gap ratio $2\Delta_{s1}(0)/k_BT_c$ is found to be 7.61, belonging to a family of strongly coupled superconductors. Our findings provide valuable insights into the superconducting gap structures of the pressure-induced superconducting nickelates.

**Introduction**

The discovery of quasi-two-dimensional layered cuprate superconductors has ignited a surge of interest in exploring novel high-temperature superconductors and elucidating their underlying mechanisms. In the cuprate system, doping a Mott insulator with strong correlation effects introduces hole carriers, and metallizes the Cu-O σ-bonds within the layers and S = 1/2 spin state, leading to the high-temperature superconductivity[1-3]. The strong hybridization between the Cu-$3d_{x^2-y^2}$ and O-$2p$ orbitals leads to the formation of Zhang-Rice singlets, which results in $d$-wave superconducting pairing. For Fe-based superconductors, the multiple $3d$ bands near Fermi surfaces contribute to $s$-wave superconducting symmetry[4]. Nickelates, possessing structural features similar to cuprate superconductors, also exhibit multi-orbital characteristics akin to those of Fe-based superconductors, have emerged as candidates for high-temperature superconductors[5-8]. The important progress in nickelates begins with the pressure-induced superconductivity in bilayer $La_3Ni_2O_{7-\delta}$ with a critical temperature $T_c$ near 80 K[9-16]. Reminiscent of the $CuO_6$ octahedra in cuprates, the doubly stacked $NiO_6$ octahedra are the key structure for the superconductivity. One of the important questions that needs to be addressed in this system concerns the symmetry and magnitude of the superconducting pairing gap. There are already many theoretical predictions on this topic, but concurrently, theoretical studies have not yet converged on a definitive conclusion[17-23]. The numerous disputes between theoretical studies and the absence of experimental results underscore the necessity for further experimental explorations to approach the underlying superconducting pairing mechanism in pressure-induced high-temperature superconductor $La_3Ni_2O_7$. Thus, investigating the magnitude and structure of the superconducting gap at high pressure is crucial and genuine for unraveling the superconductivity mechanism of the nickelate superconductors. However, the experimental challenges imposed by high-pressure conditions make it difficult to directly realize the in-situ measurements of the superconducting gap in nickelate superconductors.

The point-contact spectroscopy (PCS), including tunneling spectra and Andreev reflection spectra, is a powerful technique for directly probing the superconducting gap structure under high pressure, and has been successfully applied in megabar pressure range[24, 25]. For example, the PCS has revealed the gap characteristics of both conventional superconductors, such as $MgB_2$[26], and unconventional high-temperature superconductors, like $Bi_2Sr_2CaCu_2O_{8+\delta}$[27]. In high-pressure studies, the PCS has confirmed the unique superconducting gap features of elemental superconductors[24, 25].

In this work, we employ the Andreev reflection spectroscopy (ARS) technique to nickelate superconductor $La_3Ni_2O_{7-\delta}$ within diamond anvil cells (DACs) to address its magnitude and structure of superconducting gaps. By matching the experimental data with the Blonder-Tinkham-Klapwijk (BTK) model[28], we determine a two-gap superconducting gap structure with the dominant *s*-wave symmetry. The present work provides essential information for investigating the superconducting gap features of the $La_3Ni_2O_{7-\delta}$ samples.

## Results

### Electrical transport measurements

To ensure accurate confirmation of zero resistance in the $La_3Ni_2O_{7-\delta}$ sample, the choice of pressure-transmitting medium (PTM) is crucial. The preliminary experimental measurements indicate that the ammonia borane ($NH_3BH_3$) is chosen as PTM, which can guarantee the hydrostatic pressure condition and zero resistance in the whole pressure range. Thus, we loaded the colloid $NH_3BH_3$ into the sample chamber as the PTM to conduct all the experimental investigations. As shown in Fig. 1a, four DACs with electrodes are employed for electrical transport measurements. In Cell 1, a resistance drop is observed at 70 K, which is consistent with previous work[9, 13, 29]. In the temperature range of 70-140 K, the *R-T* curve exhibits a linear tendency, which is characteristic of the strange metal behavior[13]. To confirm the reproducibility of the zero resistance, another three high-pressure DACs are also conducted. Zero resistance is detected on these four samples (Fig. 1b), which is vital for the following ARS measurements. Besides, as reported in a recent work, the sample is easily influenced by the oxygen vacancies[30], so it's not surprising to reach a much broader superconducting state, with the zero resistance below 6 K, and onset $T_c$~72 K at 23 GPa in Cell 2. Fig. 1c illustrates the temperature dependence of the resistance of Cell 1 under applied magnetic fields ranging from 0-8 T at 23 GPa. Under external magnetic fields, the superconducting transitions are broadened, and the zero resistance is suppressed. Comparative analysis reveals that the single-band model fails to adequately describe the upper critical field $\mu_0H_c(0)$, whereas the two-band model provides a more accurate fitting. By using the two-band model[31], we extrapolate the $\mu_0H_c(0)$ as 97.7 T. The coherent lengths, $\xi^{two\ band}$ is estimated to be 18.3 Å, via the equation:

$$\mu_0H_{c2}(0) = \Phi_0/2\pi\xi^2 \qquad (1)$$

These values are also consistent with the previous report[13]. The magnitudes of these coherence lengths are well below typical values for type-I superconductors, indicating that this superconducting phase belongs to a type-II superconductor.

During the loading of $La_3Ni_2O_{7-\delta}$, we found that the probability of detecting superconducting signals was only about 20% despite in high-quality single crystal sample. Even though the superconducting volume fraction was low and some oxygen vacancies were present, it is puzzling that this issue occurs in samples larger than 30 μm. Additionally, we discovered that the samples were highly sensitive to hydrostatic pressure and exhibited anisotropy. We could only achieve the zero-resistance state under optimal hydrostatic pressure conditions. By using $NH_3BH_3$ as the PTM, we observed a zero-resistance signal at temperatures below 32 K. This result is comparable to that reported by Zhang et al.,[13] who used Daphne 7373 and observed a similar zero-resistance temperature. Therefore, $NH_3BH_3$ provides a reasonably good hydrostatic pressure environment, which lays a solid foundation for the subsequent ARS measurements.

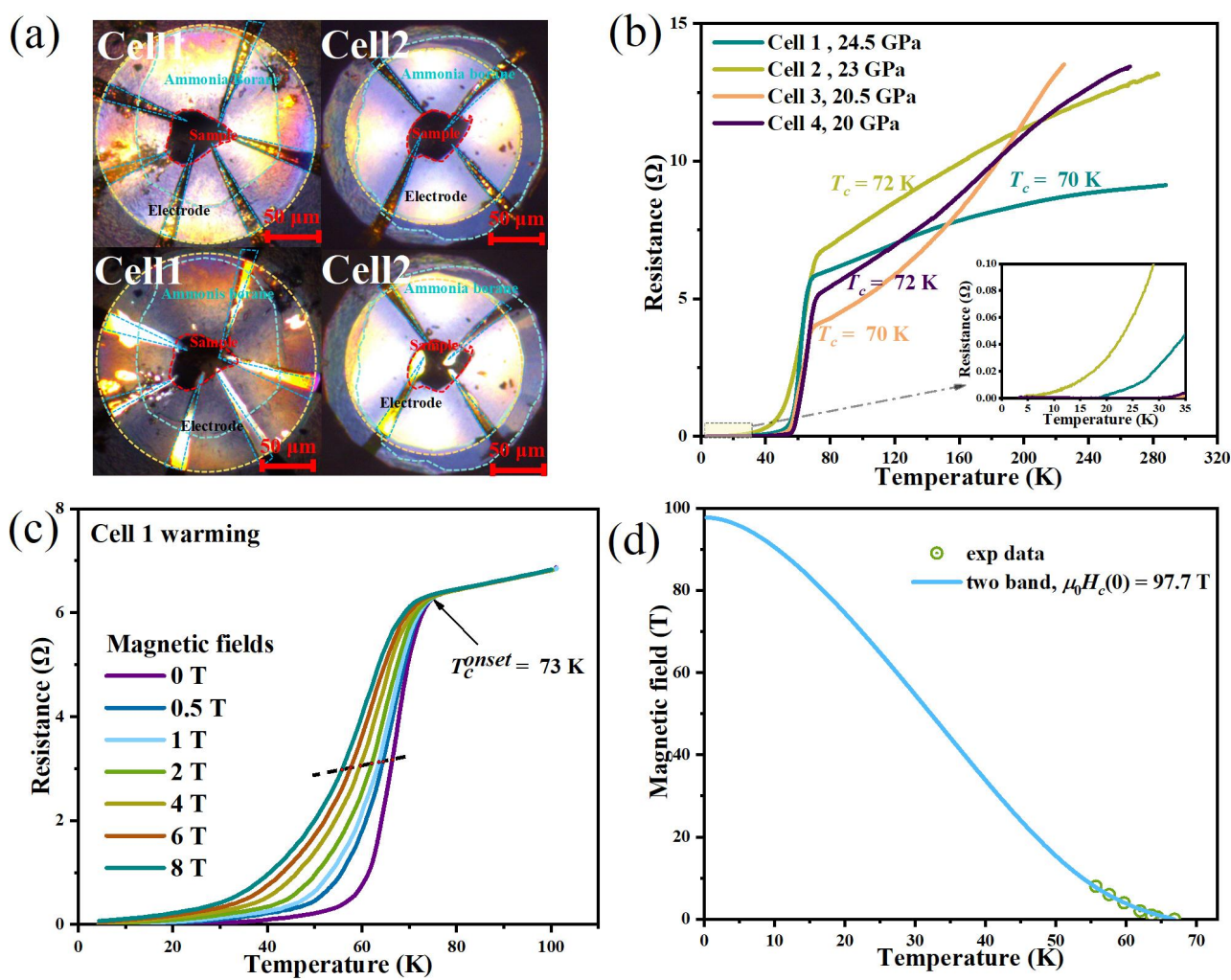

**Figure 1. The sample chamber and temperature-dependent resistance of $La_3Ni_2O_{7-\delta}$ under different pressures and external magnetic fields.** (a) Optical micrographs of the Cell 1 and Cell 2. (b) The temperature-dependence of resistance for Cell 1, Cell 2, Cell 3, and Cell 4 at 24.5 GPa and 23 GPa, 20.5 GPa and 20 GPa, respectively. The inset shows the temperatures at which zero

resistance is observed in both pressure cells. (c) The shift of the *R-T* curve under external magnetic fields ranging from 0 to 8 T at 23 GPa in Cell 1. (d) The upper critical magnetic fields of the sample, fitted using the two-band model[31].

**Andreev reflection spectra measured under high pressure**

Based on the principle of the ARS technique, we have developed a system for high-pressure ARS measurements. To test the reliability of the system, we have performed the measurements on a standard Nb metal sample. After matching the data using the BTK model, we find that the energy gap $\Delta$= 1.5 meV, in good consistent with previously reported measurements [32-34]. The matching parameters have been provided in the Extended Data Fig.1. To detect and characterize the superconducting gap features of $La_3Ni_2O_{7-\delta}$, we have performed the differential conductance measurements as a function of temperature across the junctions (Extended Data Fig.2 to Fig. 5). We pre-compressed the tip of the 3-μm thick Pt and Au electrodes to 1 μm to prevent excessive extension of the electrodes under high pressure. This has guaranteed that the contact radius between the electrode and the sample was below 3 μm, which was smaller than the maximum probe size used in conventional methods[25]. Although the contact area is significantly larger than the electron mean free path in the sample, we have observed reproducible differential conductance curves that contain superconducting gap features.

Previous experiments have found that the superconductivity of the samples was highly sensitive to oxygen vacancies and hydrostatic pressure conditions[9, 13, 29]. Therefore, we hypothesize that fragmentation occurs at the interface between the $La_3Ni_2O_{7-\delta}$ single crystal and the electrodes under high pressure. It leads to the disruption of superconductivity in small regions and forms a number of tiny superconducting channels. These channels form point-contact-like junctions with the metallic electrodes (shown in Fig. 2a), which helps us to obtain the ARS signal of $La_3Ni_2O_{7-\delta}$. Combining literature data with our experimental results, we estimated the contact area of the point-contact junctions. Assuming an average Fermi momentum of $0.5\pi/a_0$ (where the in-plane lattice constant $a_0$ = 3.7 Å)[15], a normal state resistivity of $9*10^{-6} \Omega \cdot m$, and a charge carrier density of $1.25*10^{27}$ m$^{-3}$ [35], the estimated electron mean free path is approximately 1.6 nm (The electron mean free path can be estimated by $l = \frac{m_e v_F}{ne^2\rho}$, which $m_e$ is electronic mass, $v_F$ is the Fermi velocity. The *n*, *e* and $\rho$ are carrier density, electron charge and electrical resistivity, respectively.) Using the Sharvin formula ($Rs = \frac{4\rho l}{3\pi a^2}$, the a is contact radius)[36], we calculated a contact radius a of about 7.33 nm, which is roughly five times the mean free path. This places the contact regime between the quasi-ballistic and quasi-diffusive limits. In this case, the measured typical normalized differential conductance curves exhibit two superconducting gap features, with gap peaks of approximately 22%. This phenomenon may be attributed to the low superconducting volume fraction in the sample[9, 37], causing partial conduction to occur in non-superconducting regions.

We select the results collected from Cell 2 to analyze, which provide a more stable signal for subsequent experimental data acquisition. Given the possible influence of thermal smearing and inelastic scattering effects, we consistently utilize the BTK model[28] to match the original data. The BTK model is widely recognized as a reliable method for describing the transport properties of superconducting quantum ARS junctions, which incorporates three parameters: the energy gap value $\Delta$, the junction barrier strength parameter Z, and the quasiparticle lifetime broadening parameter $\Gamma$. We compared our experimental data with different superconducting symmetry models to obtain specific superconducting gap information (Figs. 2b to 2e). In Fig. 2b, we have matched the experimental data with the two-gap *s*-wave model, and summarized the extracted superconducting gap as a function of temperature at 23 GPa. The observed larger gap value of $2\Delta_{s1}(0) \approx 46$ meV corresponds to a ratio of $2\Delta_{s1}(0)/k_BT_c \approx 7.61$. This finding confirms that $La_3Ni_2O_{7-\delta}$ is a strongly coupled superconductor. In addition to the coherence peak observed at 23 meV and 6 meV, a side peak at around 60 meV (3 K) could also be traced, which can be attributed to the collective bosonic mode[38]. This peak is typically attributed to the interaction between quasiparticles and bosonic collective excitations or the lattice vibrations. Both the bosonic mode peak and the superconducting coherence peak weaken and even disappear simultaneously with increasing temperature, exhibiting a connection between the bosonic mode and the superconducting state. The hump-like feature curves have also been observed in various unconventional superconductor systems like cuprate $Bi_2Sr_2CaCu_2O_{8+\delta}$[39], nickelate $Nb_{1-x}Sr_xNiO_2$[38], Fe-based superconductors[40], the Kagome superconductor $CsV_{3-x}Ta_xSb_5$[41], and the thin film $La_2PrNi_2O_7$[42].

To determine the superconducting pairing symmetry, we compared the experimental results with some other different models (*s* + *d*-wave, *d* + *s*-wave, and two *d*-wave, see Figs. 2c to 2e). Subsequent matchings using an *s*-wave for the larger gap and a *d*-wave for the smaller gap also reproduced the experimental data well, and the extracted gap values remained nearly unchanged. However, when the dominant gap is set to *d*-wave, whether in the *d* + *d* or *d* + *s* configuration, the theory model failed to match the experimental data. Although our results support an *s*-wave dominant two-gap feature, a definitive conclusion regarding the pairing symmetry still requires more experimental evidence.

To ensure the reliability of the signals, we performed ARS on three other $La_3Ni_2O_{7-\delta}$ samples as well as the temperature-dependent measurements on all the diamond anvil cells (DACs) (see Extended Data Figs. 2 to 5). As shown in Extended Data Fig.11, at 3 K, the gap features carried by the curves were generally consistent across the samples. In addition, we performed normalization on the low-temperature differential conductance spectra from Cell 1 and Cell 2 and matched them using a two-band *s*-wave model (see Extended Data Figs. 9 and 10). By comparing the spectra from different cells, we observed kink-like features near the superconducting gap and bosonic mode energies in all DACs. Cells 3 and 4 also exhibited the signs of two-gap behavior. The ZBCP can be observed at different positions in Cell 2, which appears only at high temperatures. We attribute this to a reduced barrier parameter Z with increased temperatures (shown in Extended Data Fig.10), as similarly discussed in [43], which in turn made the smaller gap more apparent.

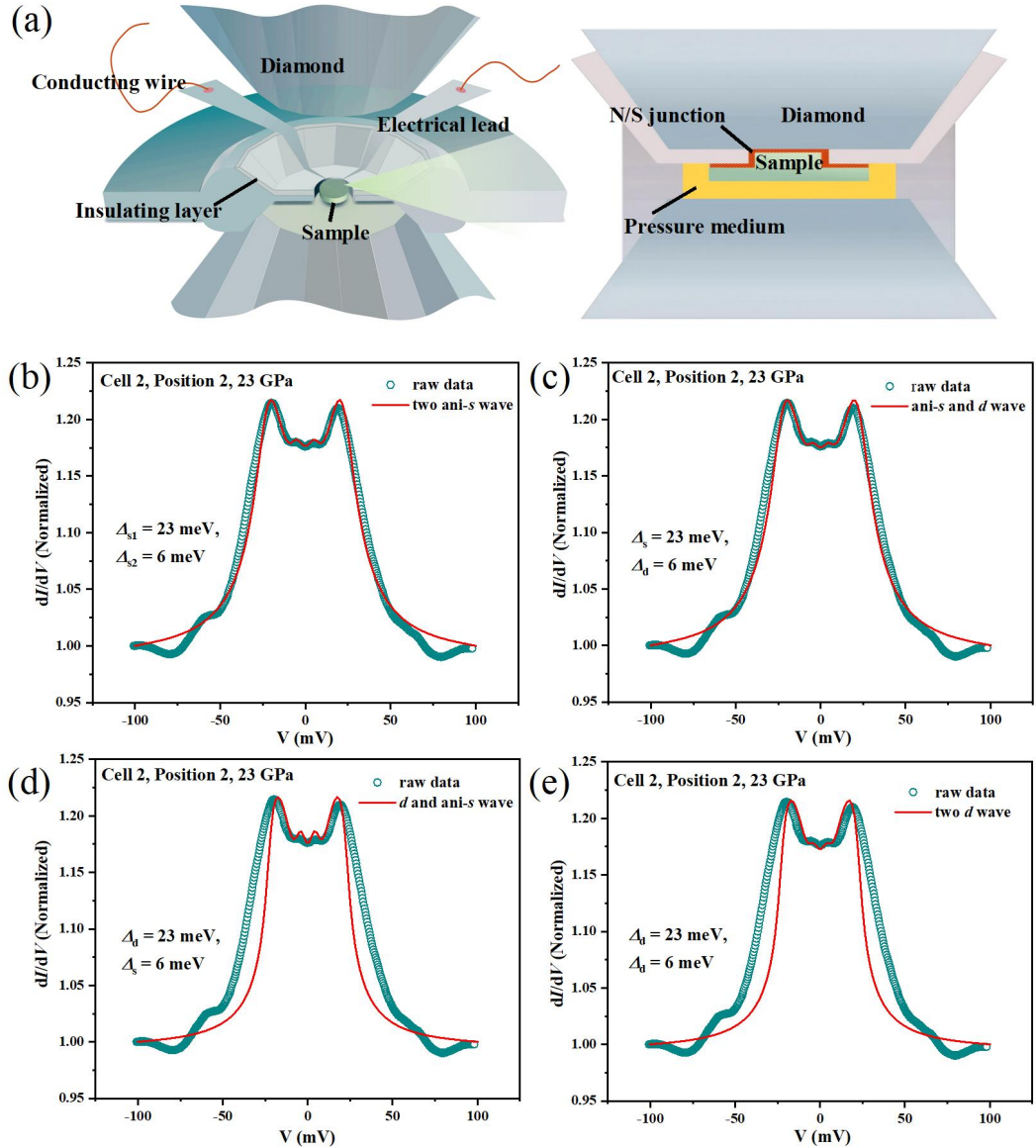

**Figure 2. Andreev reflection spectra matched using the BTK model at different temperatures.** (a) Schematic diagram of the setup used for high-pressure ARS measurements. (b), (c), (d), and (e) The differential conductance curve of Cell 2 (Position 2 at 3 K) matching by different *s+s*, *s+d*, *d+s*, *d+d* models, respectively.

Figure. 3a illustrates the normalized conductance curves with a two-gap structure under different temperatures at 23 GPa. These spectra have been normalized by the polynomial fitting model[44] (where T is temperature and V is the bias voltage). In principle, we can determine the superconducting gap from the inflection points of the conductance curve; however, the value may be overestimated due to thermal effects and signal quality[24, 45, 46]. At the lowest measured temperature of 3 K, the coherence peaks of the conductance curve are observed at 23 and 6 meV, respectively. Along with the superconducting gap decreasing with increasing temperatures, the differential conductance peak width gradually narrows. As shown in the first derivative of the R-T curve from Cell 2 (see

Extended Data Fig.6), the sample resistance increases sharply above 40 K. The thermal broadening effect causes the differential conductance curves to lose clear two-gap features above 40 K (see Extended Data Fig.7), making it difficult to obtain the intrinsic characteristics of the superconducting gaps in high-temperature region. Figure 3b summrizes the temperature dependence of superconducting gap values, which are matched by a two-gap s-wave model. The evolution of both gaps with temperature follows the BTK model, and the extrapolated critical temperature is 72 K, which is consistent with the results from transport measurements.

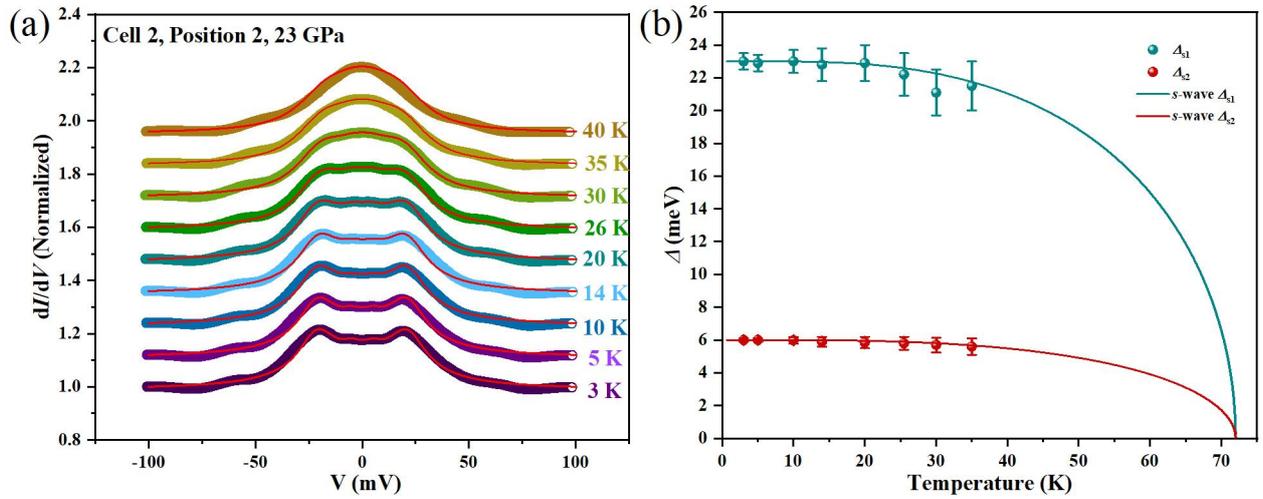

**Figure 3. Andreev reflection spectra (ARS) matched using the BTK model at different temperatures.** (a) The conductance curve was normalized by the polynomial curve, and the two s-wave BTK model was used to match the ARS at different temperatures (in the temperature range of 3-40 K). (b) The superconducting gap values obtained from the BTK fitting at 23 GPa.

**The critical current of La$_3$Ni$_2$O$_{7-\delta}$ single crystal**

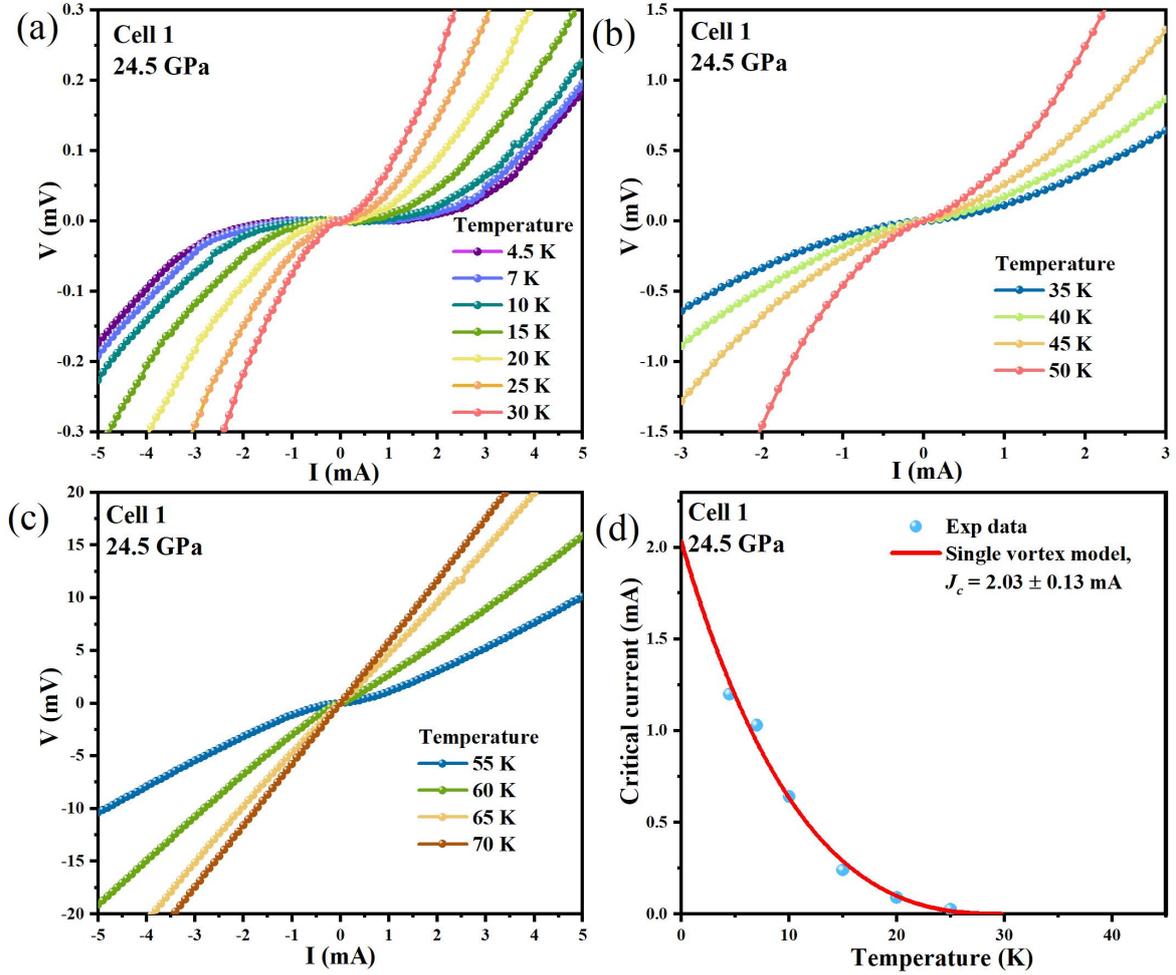

**Figure 4. *I-V* curves of La$_3$Ni$_2$O$_{7-\delta}$ sample at different temperatures.** (a), (b), and (c) show the *I-V* curves in the temperature ranges below 30 K, between 35 K and 50 K, and above 50 K, respectively. (d) The critical current obtained by fitting with the single vortex model.

To eliminate the potential influence of critical current ($J_c$) on the inflection points observed in the differential conductance, we have measured the critical current of the sample in the range of -5 to 5 mA below the $T_c$ (Fig. 4). In the *I-V* curve, the current value at which the voltage becomes non-zero is defined as the $J_c$. Above the temperature of 30 K, the voltage remains non-zero across the measurement current range. At 70 K, the *I-V* curve exhibits a distinct linear behavior. We fit the temperature dependence of the critical current using the vortex pinning model[47] and obtained $J_c(0)$ = 2.03 mA. We estimate the cross-section area of the sample to be 50 μm × 5 μm, leading to a derived critical current density of 812 A/cm$^2$, which is consistent with the reported result[13]. The relatively small critical current is possibly attributed to the low superconducting volume fraction and inhomogeneity of the sample[37, 48]. Importantly, this critical current value exceeds the current used in the inflection points for the differential conductance, effectively excluding the possibility that excessive current and resultant thermal effects could affect the present spectroscopy measurements.

**Discussions**

Recently, the theoretical calculations have proposed a broad spectrum of possibilities for the superconducting pairing symmetry in La$_3$Ni$_2$O$_{7-\delta}$[20, 49-57], but a unified conclusion on the pairing mechanism is still lacking. As a pressure-induced nickelate high-temperature superconductor, bilayer La$_3$Ni$_2$O$_{7-\delta}$ exhibits structural characteristics distinct from both cuprate superconductors and infinite-layer nickelates. In La$_3$Ni$_2$O$_{7-\delta}$, in addition to the in-plane $d_{x^2-y^2}$ orbitals of the Ni-O plane, the $d_{3z^2-r^2}$ orbitals in the interlayer direction also play a significant role. According to current theoretical models, the in-plane $d_{x^2-y^2}$ orbitals lead to $d$-wave superconducting symmetry in La$_3$Ni$_2$O$_{7-\delta}$, similar to cuprates and infinite-layer nickelates[20, 49, 56, 57], while interlayer coupling mediated by the $d_{3z^2-r^2}$ orbitals induces $s\pm$-wave pairing[50-55].

Recently, Liu et al. conducted ARS on La$_3$Ni$_2$O$_{7-\delta}$ single crystals under high pressure[35]. They observed a zero-bias conductance peak (ZBCP) and multiple-gap features. By fitting the experimental data, they attributed the ZBCP to $d$-wave symmetry, while the other two larger gaps were considered as $s$-wave. However, the two larger gaps, which were determined based on the kinks in the differential conductance, were suggested to derive from some other sources. Additionally, superconductivity at about 40 K under ambient pressure was recently achieved in La$_2$PrNi$_2$O$_{7-\delta}$ thin films. Fan et al. performed scanning tunneling spectroscopy on these films and observed two-gap features and a bosonic mode[42]. By fitting the data with the Dynes model, they thought the two-gap $d$-wave couldn't fit experimental data. The spectra could be well fitted using two-gap anisotropic $s$-wave or an anisotropic $s + d$-wave pairing, yielding gap values of $\Delta_1$ = 19 meV and $\Delta_2$ = 6 meV. Considering that the $T_c$ of La$_2$PrNi$_2$O$_{7-\delta}$ thin film is lower than the bulk La$_3$Ni$_2$O$_{7-\delta}$, although their gap values are lower, their analysis results on symmetry are consistent with this work.

A series of angle-resolved photoemission spectroscopy (ARPES) experimental studies under ambient pressure have offered further insights into the superconducting pairing mechanism in

Ruddlesden–Popper phase nickelates. The ARPES measurements about $La_3Ni_2O_{7-\delta}$ reported a flat band associated with the Ni-$3d_z^2$ orbitals at approximately 50 meV below the Fermi level[10]. The subsequent theoretical calculations predicted that this flat band crossed the Fermi surface under pressure, and pointed to the $s\pm$-wave pairing model. The recent discovery of ambient pressure superconductivity in nickelate thin films brings new opportunities for probing the pairing mechanism in these materials. Notably, Li et al. and Wang et al. both performed ARPES on $La_{3-x}Pr_xNi_2O_{7-\delta}$ films and confirmed the contribution of the Ni $d_{x2-y2}$ orbital at the Fermi level [58, 59]. In our study, through a comparative analysis of experimental spectra and theoretical models, we preliminarily identified a $s$-wave dominant two-gap superconducting feature. The future advancements in high-quality single-crystal growth and high-pressure tunneling spectra experiments may offer new insight into this issue.

**Conclusion**

In this work, we in situ fabricate conductance junctions within DACs and conduct Andreev reflection spectroscopy measurements on $La_3Ni_2O_{7-\delta}$ samples exceeding 20 GPa. We have used $NH_3BH_3$ as the PTM and successfully achieved zero resistance measurements below 32 K. By matching the differential conductance curves of $La_3Ni_2O_{7-\delta}$ at various temperatures with the BTK model, we determine a two-gap structure with superconducting gap values $2\Delta$ of 46 meV and 12 meV, with the superconducting symmetry predominantly exhibiting $s$-wave character. The $T_c$ determined from the fitted superconducting gap is 72 K, in agreement with the results from the electrical transport measurements. This work addresses a crucial insight into the understanding of the superconducting gap characteristics of novel high-temperature nickel-based superconductors.

**Methods**

The La$_3$Ni$_2$O$_{7-\delta}$ sample was grown using the same method as reported in the previous work[9]. Four symmetric diamond anvil cells (DACs) with 200 and 300 μm culet anvils were applied for electrical transport measurements and Andreev reflection spectroscopy (ARS). The pressure was determined by measuring the Raman edge of the diamond using Akahama's calibration[60]. The indentation of the tungsten gasket was insulated with c-BN/epoxy, and the bevel area was filled with MgO/epoxy. Ammonia borane (NH$_3$BH$_3$) was employed as the pressure-transmitting medium. To optimize the ARS signals, six electrodes were integrated into the diamond by manually cutting platinum (Pt) foil with a thickness of 3 μm (though only four electrodes were used for measurements). Another three DACs with 200 μm culet anvils were also utilized for resistance measurements and PCS. The preparation and pressure calibration methods for this DAC were identical to the first one.

According to the R-T curves measured under external magnetic fields, the upper critical field of La$_3$Ni$_2$O$_{7-\delta}$ far exceeds the 9 T field limit of the commercial magnet available in our laboratory (9T). Therefore, we were unable to suppress the superconducting transition by applying an external magnetic field at the current temperature range, which prevented us from obtaining a normalized background curve under the normal state. And during variable-temperature measurements, the differential conductance curves exhibited distortions at high temperatures. Besides, the additional resistance brought by the three-electrode measurements also leads to the asymmetry of differential conductance curves. As a result, we employed polynomial fitting to normalize the differential conductance spectra.

The van der Pauw method was employed to measure electrical resistance in a helium cryostat (1.5 K–300 K) equipped with a 0–9 T superconducting magnet. We used the delta model of the Keithley current source (Model 6221, 0.2 mA) and a Keithley voltmeter (Model 2182A) to measure sample resistance.

For differential conductance d$I$/d$V$ (V) measurements, a STANFORD Lock-in Amplifier (SR830) was used to supply a small AC current. To ensure the current magnitude remained around the microampere range, a 48 kΩ resistor was connected to the current output of the lock-in amplifier. In order to ensure the accuracy of the test system, the Nb foil was used as the standard sample. By fitting the differential conductance curve of Nb at 3 K, suggests the superconducting gap of 2$\Delta$ = 3 meV, which is consistent with consistent with the previous conclusion ( as shown in Extended Data Fig.1).

**Data availability**

All data are available from the corresponding authors upon reasonable request. Source data are provided with this paper.

**Acknowledgements**

This work was supported by National Key R&D Program of China (2022YFA1405500), the National Natural Science Foundation of China (52372257, 12425404, 12474137, 12494459010), the National Key Research and Development Program of China (2023YFA1406000), the Guangdong Basic and Applied Basic Research Foundation (2024B1515020040, 2025B1515020008 and 2024A1515030030), the Shenzhen Science and Technology Program (RCYX20231211090245050), the Guangdong Provincial Key Laboratory of Magnetoelectric Physics and Devices (2022B1212010008), and the Research Center for Magnetoelectric Physics of Guangdong Province (2024B0303390001). The authors express thankfulness to Prof. Huan Yang, Dr. Wuhao Chen and Dr. Jinyulin- Li for intensive discussions.

**Author contributions**

X. Huang conceived this project. JN. Guo, YZ. Chen, YL. Wang, and X. Huang carried out the instrument calibration, ARS and transport property measurements. M. Wang, HL. Sun, and D Hu performed the sample synthesis. J. Guo, Y. Chen and X. Huang wrote the paper with contributions from all authors.

**Competing interests**

The authors declare no competing interests.

**References**

1. Zhang FC, Rice TM. Effective Hamiltonian for the superconducting Cu oxides. *Physical Review B* **37**, 3759-3761 (1988).
2. Lee PA, Nagaosa N, Wen X-G. Doping a Mott insulator: Physics of high-temperature superconductivity. *Reviews of Modern Physics* **78**, 17-85 (2006).
3. Bednorz JG, Müller KA. Possible highTc superconductivity in the Ba−La−Cu−O system.

*Zeitschrift für Physik B Condensed Matter* **64**, 189-193 (1986).

4. Chen X, Dai P, Feng D, Xiang T, Zhang F-C. Iron-based high transition temperature superconductors. *National Science Review* **1**, 371-395 (2014).

5. Li D, *et al.* Superconductivity in an infinite-layer nickelate. *Nature* **572**, 624-627 (2019).

6. Osada M, Wang BY, Lee K, Li D, Hwang HY. Phase diagram of infinite layer praseodymium nickelate $Pr_{1-x}Sr_xNiO_2$ thin films. *Physical Review Materials* **4**, 121801 (2020).

7. Pan GA, *et al.* Superconductivity in a quintuple-layer square-planar nickelate. *Nature Materials* **21**, 160-164 (2022).

8. Wang NN, *et al.* Pressure-induced monotonic enhancement of Tc to over 30 K in superconducting $Pr_{0.82}Sr_{0.18}NiO_2$ thin films. *Nature Communications* **13**, 4367 (2022).

9. Sun H, *et al.* Signatures of superconductivity near 80 K in a nickelate under high pressure. *Nature* **621**, 493-498 (2023).

10. Yang J, *et al.* Orbital-dependent electron correlation in double-layer nickelate $La_3Ni_2O_7$. *Nature Communications* **15**, 4373 (2024).

11. Hou J, *et al.* Emergence of High-Temperature Superconducting Phase in Pressurized $La_3Ni_2O_7$ Crystals. *Chin Phys Lett* **40**, (2023).

12. Liu Z, *et al.* Electronic correlations and partial gap in the bilayer nickelate La3Ni2O7. *Nature Communications* **15**, 7570 (2024).

13. Zhang Y, *et al.* High-temperature superconductivity with zero resistance and strange-metal behaviour in $La_3Ni_2O_{7-\delta}$. *Nature Physics* **20**, 1269-1273 (2024).

14. Wang G, *et al.* Pressure-Induced Superconductivity In Polycrystalline $La_3Ni_2O_{7-\delta}$. *Physical Review X* **14**, 011040 (2024).

15. Wang L, *et al.* Structure Responsible for the Superconducting State in $La_3Ni_2O_7$ at High-Pressure and Low-Temperature Conditions. *Journal of the American Chemical Society* **146**, 7506-7514 (2024).

16. Li J, *et al.* Identification of Superconductivity in Bilayer Nickelate La3Ni2O7 under High Pressure up to 100 GPa. *National Science Review*, nwaf220 (2025).

17. Lu C, Pan Z, Yang F, Wu C. Interlayer-Coupling-Driven High-Temperature Superconductivity in L3Ni2O7 under Pressure. *Physical Review Letters* **132**, 146002 (2024).

18. Jiang R, Hou J, Fan Z, Lang Z-J, Ku W. Pressure Driven Fractionalization of Ionic Spins Results in Cupratelike High-$T_c$ Superconductivity in $La_3Ni_2O_7$. *Physical Review Letters* **132**, 126503 (2024).

19. Fan Z, *et al.* Superconductivity in nickelate and cuprate superconductors with strong bilayer coupling. *Physical Review B* **110**, 024514 (2024).

20. Jiang K, Wang Z, Zhang F-C. High-Temperature Superconductivity in $La_3Ni_2O_7$. *Chinese Physics Letters* **41**, 017402 (2024).

21. Tian Y-H, Chen Y, Wang J-M, He R-Q, Lu Z-Y. Correlation effects and concomitant two-orbital s±-wave superconductivity in La$_3$Ni$_2$O$_7$ under high pressure. *Physical Review B* **109**, 044801 (2024).

22. Liao Z, *et al.* Electron correlations and superconductivity in La$_3$Ni$_2$O$_7$ under pressure tuning. *Physical Review B* **108**, 214522 (2023).

23. Yang Y-f, Zhang G-M, Zhang F-C. Interlayer valence bonds and two-component theory for high-Tc superconductivity of La$_3$Ni$_2$O$_7$ under pressure. *Physical Review B* **108**, L201108 (2023).

24. Du F, *et al.* Tunneling Spectroscopy at Megabar Pressures: Determination of the Superconducting Gap in Sulfur. *Physical Review Letters* **133**, 036002 (2024).

25. Cao Z-Y, *et al.* Spectroscopic evidence for the superconductivity of elemental metal Y under pressure. *NPG Asia Materials* **15**, 5 (2023).

26. Daghero D, *et al.* Point-contact spectroscopy in MgB$_2$ single crystals in magnetic field. *Physica C: Superconductivity* **385**, 255-263 (2003).

27. Gu Q, *et al.* Directly visualizing the sign change of d-wave superconducting gap in Bi$_2$Sr$_2$CaCu$_2$O$_{8+\delta}$ by phase-referenced quasiparticle interference. *Nature Communications* **10**, 1603 (2019).

28. Blonder GE, Tinkham M, Klapwijk TM. Transition from metallic to tunneling regimes in superconducting microconstrictions: Excess current, charge imbalance, and supercurrent conversion. *Physical Review B* **25**, 4515-4532 (1982).

29. Wang G, *et al.* Pressure-Induced Superconductivity In Polycrystalline La$_3$Ni$_2$O$_{7-\delta}$. *Physical Review X* **14**, 011040 (2024).

30. Dong Z, *et al.* Visualization of oxygen vacancies and self-doped ligand holes in La(3)Ni(2)O$_{(7\text{-delta})}$. *Nature* **630**, 847-852 (2024).

31. Gurevich A. Enhancement of the upper critical field by nonmagnetic impurities in dirty two-gap superconductors. *Physical Review B* **67**, 184515 (2003).

32. Hafstrom JW, Rose RM, Macvicar MLA. Evidence for a second energy gap in superconducting niobium. *Physics Letters A* **30**, 379-380 (1969).

33. Janson L, Klein M, Lewis H, Lucas A, Marantan A, Luna K. Undergraduate experiment in superconductor point-contact spectroscopy with a Nb/Au junction. *American Journal of Physics* **80**, 133-140 (2012).

34. Hafstrom JW, MacVicar MLA. Case for a Second Energy Gap in Superconducting Niobium. *Physical Review B* **2**, 4511-4516 (1970).

35. Liu C, *et al.* Andreev reflection in superconducting state of pressurized La$_3$Ni$_2$O$_7$. *Science China Physics, Mechanics & Astronomy* **68**, 247412 (2025).

36. Daghero D, Gonnelli RS. Probing multiband superconductivity by point-contact spectroscopy.

*Superconductor Science and Technology* **23**, 043001 (2010).

37. Yazhou Zhou, *et al.* Investigations of key issues on the reproducibility of high-Tc superconductivity emerging from compressed La3Ni2O7. *arXiv:231112361*, (2023).

38. Gu Q, *et al.* Single particle tunneling spectrum of superconducting $Nd_{1-x}Sr_xNiO_2$ thin films. *Nature Communications* **11**, 6027 (2020).

39. Lee J, *et al.* Interplay of electron–lattice interactions and superconductivity in $Bi_2Sr_2CaCu_2O_{8+\delta}$. *Nature* **442**, 546-550 (2006).

40. Wang Z, *et al.* Close relationship between superconductivity and the bosonic mode in $Ba_{0.6}K_{0.4}Fe_2As_2$ and $Na(Fe_{0.975}Co_{0.025})As$. *Nature Physics* **9**, 42-48 (2012).

41. Hu B, *et al.* Evidence of a distinct collective mode in Kagome superconductors. *Nature Communications* **15**, 6109 (2024).

42. Fan S, *et al.* Superconducting gaps revealed by STM measurements on La2PrNi2O7 thin films at ambient pressure.) (2025).

43. Samuely P, Szabó P, Pribulová Z, Tillman ME, Bud'ko SL, Canfield PC. Possible two-gap superconductivity in NdFeAsO0.9F0.1 probed by point-contact Andreev-reflection spectroscopy. *Superconductor Science and Technology* **22**, 014003 (2009).

44. Pecchio P, *et al.* Doping and critical-temperature dependence of the energy gaps in Ba(Fe1-xCox) 2As2 thin films. *Physical Review B* **88**, 174506 (2013).

45. Szabó P, *et al.* Evidence for Two Superconducting Energy Gaps in MgB2 by Point-Contact Spectroscopy. *Physical Review Letters* **87**, 137005 (2001).

46. Szabó P, Pribulová Z, Pristáš G, Bud'ko SL, Canfield PC, Samuely P. Evidence for two-gap superconductivity in $Ba_{0.55}K_{0.45}Fe_2As_2$ from directional point-contact Andreev-reflection spectroscopy. *Physical Review B* **79**, 012503 (2009).

47. Dew-Hughes D. Flux pinning mechanisms in type II superconductors. *The Philosophical Magazine: A Journal of Theoretical Experimental and Applied Physics* **30**, 293-305 (1974).

48. Xin Fu, Jing W. Cinvergence rate and uniform lipschitz estimate in periodic homogenization of high-contrast elliptic suystems. *arXiv:240411396*.

49. Lechermann F, Gondolf J, Bötzel S, Eremin IM. Electronic correlations and superconducting instability in La3Ni2O7 under high pressure. *Physical Review B* **108**, L201121 (2023).

50. Qin Q, Yang Y-f. High-Tc superconductivity by mobilizing local spin singlets and possible route to higher Tc in pressurized La3Ni2O7. *Physical Review B* **108**, L140504 (2023).

51. Qu X-Z, *et al.* Bilayer $t$-$J$-$J^{\perp}$ Model and Magnetically Mediated Pairing in the Pressurized Nickelate $La_3Ni_2O_7$. *Physical Review Letters* **132**, 036502 (2024).

52. Tian Y-H, Chen Y, Wang J-M, He R-Q, Lu Z-Y. Correlation effects and concomitant two-orbital s±-wave superconductivity in La3Ni2O7 under high pressure. *Physical Review B* **109**, 165154 (2024).

53. Yang Q-G, Wang D, Wang Q-H. Possible s±-wave superconductivity in $La_3Ni_2O_7$. *Physical Review B* **108**, L140505 (2023).

54. Zhang Y, Lin L-F, Moreo A, Maier TA, Dagotto E. Structural phase transition, s±-wave pairing, and magnetic stripe order in bilayered superconductor $La_3Ni_2O_7$ under pressure. *Nature Communications* **15**, 2470 (2024).

55. Liu YB, Mei JW, Ye F, Chen WQ, Yang F. s+/--Wave Pairing and the Destructive Role of Apical-Oxygen Deficiencies in La3Ni2O7 under Pressure. *Phys Rev Lett* **131**, 236002 (2023).

56. Heier G, Park K, Savrasov SY. Competing dxy and s± pairing symmetries in superconducting $La_3Ni_2O_7$ LDA+FLEX calculations. *Physical Review B* **109**, 104508 (2024).

57. Xia C, Liu H, Zhou S, Chen H. Sensitive dependence of pairing symmetry on Ni-eg crystal field splitting in the nickelate superconductor $La_3Ni_2O_7$. *Nature Communications* **16**, 1054 (2025).

58. Li P, *et al.* Angle-resolved photoemission spectroscopy of superconducting $(La,Pr)_3Ni_2O_7/SrLaAlO_4$ heterostructures. *National Science Review*, nwaf205 (2025).

59. Wang BY*, et al.* Electronic structure of compressively strained thin film La2PrNi2O7.)  (2025).

60. Akahama Y, Kawamura H. Pressure calibration of diamond anvil Raman gauge to 310GPa. *Journal of Applied Physics* **100**, 043516 (2006).

# Supplementary Materials

# for

# Revealing superconducting gap in La$_3$Ni$_2$O$_{7-\delta}$ by Andreev reflection spectroscopy under high pressure

Jianning Guo[1,†], Yuzhi Chen[1,†], Yulong Wang, Hualei Sun[2], Deyuan Hu[3], Meng Wang[3,*], Xiaoli Huang[1,*] and Tian Cui[1,4]

[1] *State Key Laboratory of High Pressure and Superhard Materials, College of Physics, Jilin University, Changchun 130012, China*

[2]*School of Sciences, Sun Yat-sen University, Shenzhen 518107, China*

[3]*Center for Neutron Science and Technology, Guangdong Provincial Key Laboratory of Magnetoelectric Physics and Devices, School of Physics, Sun Yat-Sen University, Guangzhou, China.*

[4] *School of Physical Science and Technology, Ningbo University, Ningbo 315211, China*

†These authors contributed equally to this work
*Corresponding authors, emails:
huangxiaoli@jlu.edu.cn (X. Huang)
wangmeng5@mail.sysu.edu.cn (M. Wang)

**BTK model**

The BTK model is used to describe electric transport properties across the interface between a normal metal and superconductor[1]. The potential barrier from N-S junction is regarded as a delta barrier in theory. The tunneling current $I_{NS}$ as a function of bias voltage V at finite temperature can be expressed as below:

$$I_{NS} = C \int_{-\infty}^{+\infty} [f(E - eV) - f(E)][1 + A(E) - B(E)]dE$$

So the differential conductance can be derived as:

$$G_{NS} = C \int_{-\infty}^{\infty} \frac{df(E - V, T)}{dV}[1 + A(E) - B(E)]dE$$

Where $f(E, T)$ is the Fermi-Dirac distribution, and $A(E)$ and $B(E)$ are the energy-dependent probability coefficients of Andreev reflection and normal reflection at the interface, $C$ is a constant which depends on the area of the junction, on the density of states and on the Fermi velocity. When divided by the conductance of the same junction when the superconductor is in the normal state, $dI_{NN}/dV$, this gives the normalized conductance of the junction, G. In our work, due to the broad superconducting transition width and extremely high upper critical field, we employed a polynomial-fitted curve as the differential conductance of the normal state [2].

By considering the quasiparticle lifetime broaden parameter $\Gamma$ and junction barrier strength parameter Z, as well as matching the boundary conditions, [1+A(E)-B(E)] can be expressed as the product of the transparency of the barrier $\tau_N$ and $\sigma_S$.

$$\tau_N(\theta) = \frac{\cos^2\theta}{\cos^2\theta + Z^2}$$

$$\sigma_S(E, \theta) = \frac{1 + \tau_N(\theta)|\gamma_+(E, \theta)|^2 + (\tau_N(\theta) - 1)|\gamma_+(E, \theta)\gamma_-(E, \theta)|^2}{|1 + (\tau_N(\theta) - 1)\gamma_+(E, \theta)\gamma_-(E, \theta)\exp(i\varphi_d)|^2}$$

Where

$$\gamma_\pm(E, \theta) = \frac{(E + i\Gamma) - \sqrt{(E + i\Gamma)^2 - |\Delta(E, \theta)_\pm|^2}}{|\Delta(E, \theta)_\pm|}$$

and

$$\varphi_d(\theta) = -i\ln\left[\frac{\Delta(E, \theta)_+/|\Delta(E, \theta)_+|}{\Delta(E, \theta)_-/|\Delta(E, \theta)_-|}\right]$$

Here Δ is the value of superconducting gap. $\Delta(E, \theta)_+$ and $\Delta(E, \theta)_-$ represent the quasiparticle of the electronic-like and hole-like, respectively. The data is fitted by adjusting the values of each parameter (Γ, Z, Δ and T).

For *d*-wave superconductors, The superconducting gap can be described by :

$$\Delta(E,\theta)_\pm = \Delta_0 \cdot \cos(2(\theta \mp \alpha))$$

Where α is the angle between the normal vector of the interface and the x axis of the sample.

By taking into account temperature and multiband superconductivity effects, the final normalized differential conductance can be expressed as:

$$\frac{dI_{NS}}{dV}(V) = C\frac{d}{dV}\int_{-\infty}^{+\infty}[f(E-eV,T)-f(E,T)]G_t\, dE$$

Where

$$G(E) = \frac{\int_{-\frac{\pi}{2}}^{+\frac{\pi}{2}} \sigma_S(E,\theta)\tau_N(\theta)\cos\theta\, d\theta}{\int_{-\frac{\pi}{2}}^{+\frac{\pi}{2}} \tau_N(\theta)\cos\theta\, d\theta}$$

$$G_t(E) = w_1 G_1(E) + (1-w_1)G_2(E)$$

Here G(E) is the temperature-independent normalized differential conductance. $G_i(E)$ is the normalized differential conductance of the ith band and ω1 is the relative contribution of band 1 to the total conductance.

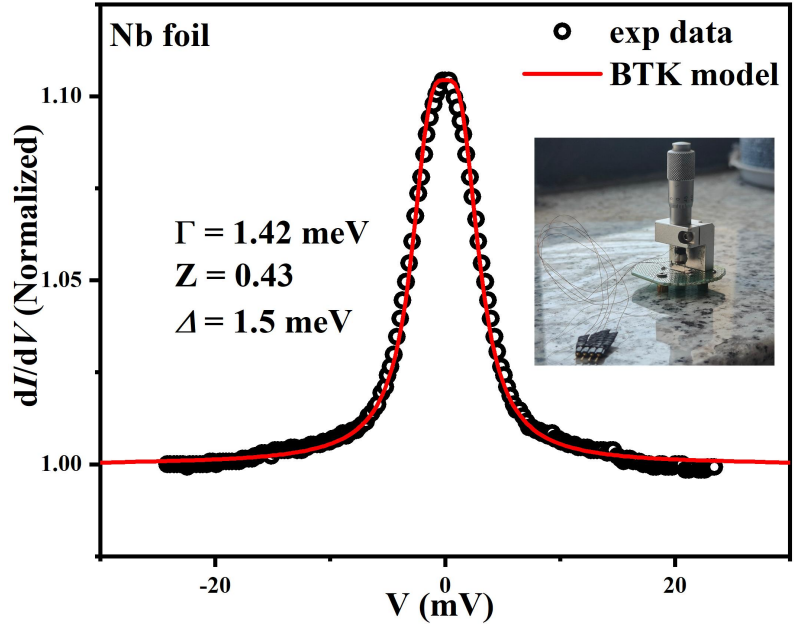

**Extended data Fig.1. The normalized differential conductance spectra of Nb foil at 3 K.** The *s*-wave BTK fit gives the superconducting gap 2*Δ*=3 meV, Z=0.43, and Γ=1.42 meV. The large Γ value introduces a low quality of the conductance spectra. The point contact measurement device has been displayed in the inset.

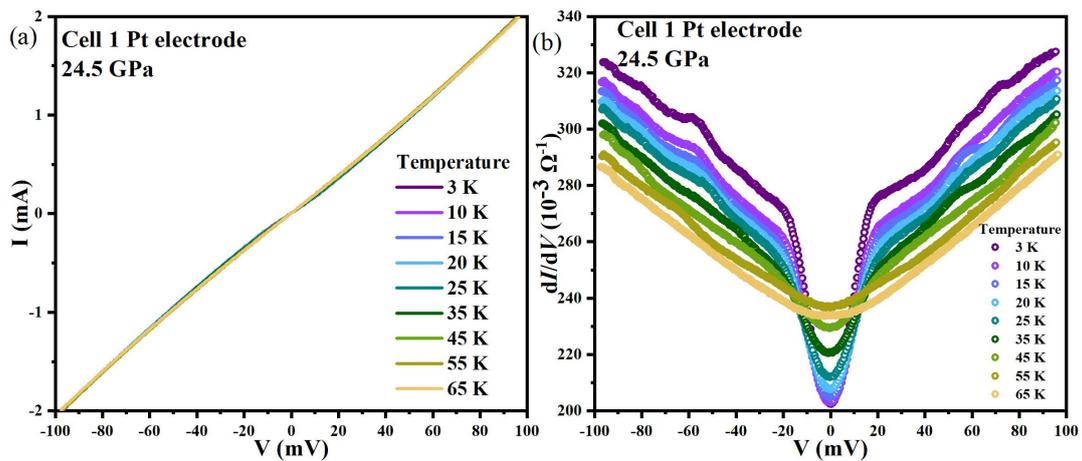

**Extended Data Fig.2. Temperature dependence of the conductance spectra.** (a) The temperature dependence of the current-voltage characteristics. (b) The raw, non-normalized differential conductance curves at various temperatures below 65 K.

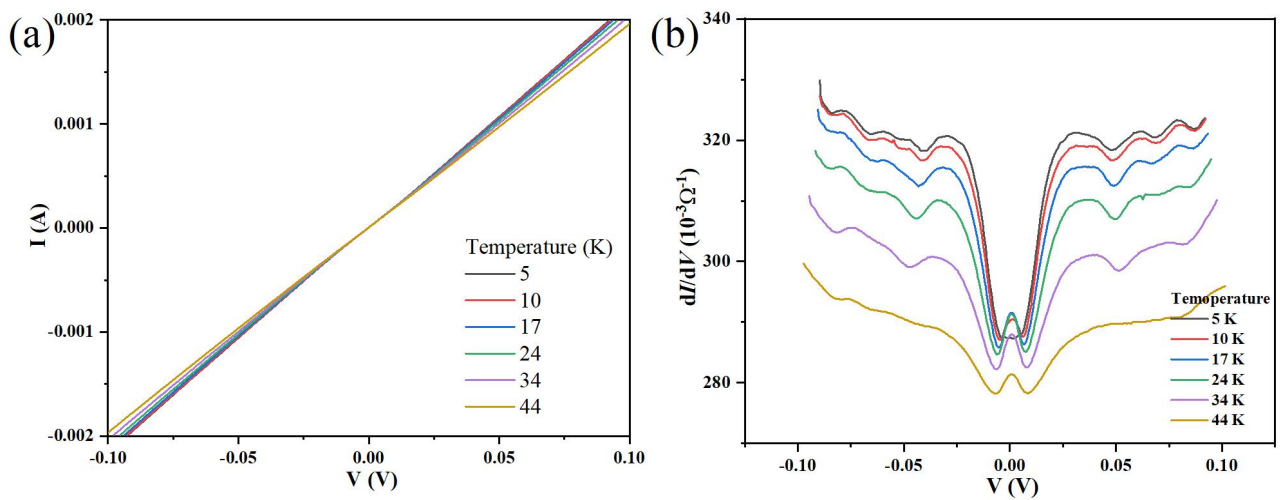

**Extended Data Fig.3. Temperature dependence of the conductance spectra of Cell 2, Position 1.** (a) The temperature dependence of the current-voltage characteristics. (b) The raw, non-normalized differential conductance curves at various temperatures below 44K.

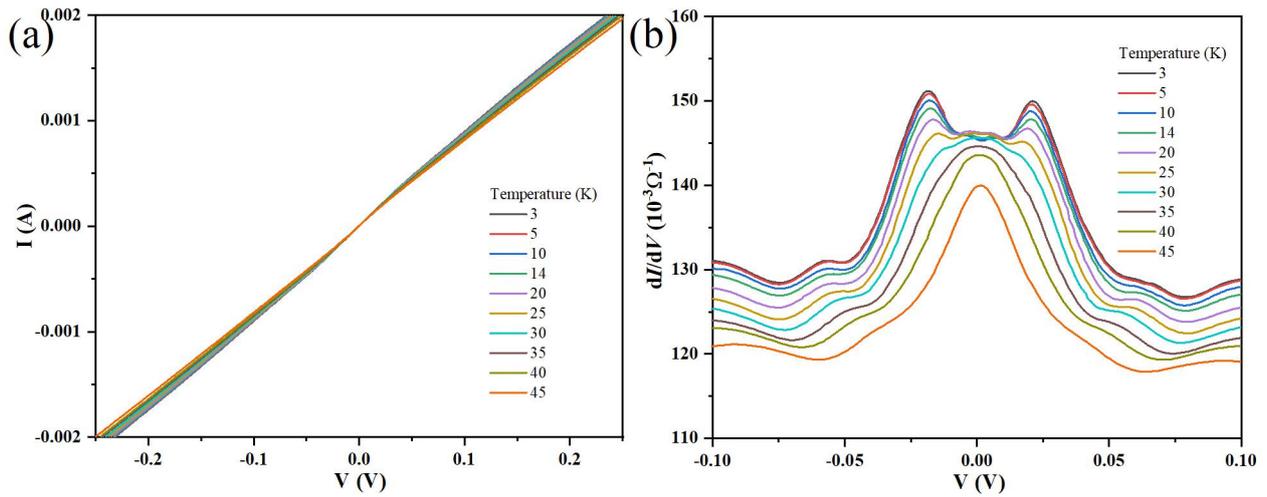

**Extended Data Fig.4. Temperature dependence of the conductance spectra of Cell 2, Position 2.**
(a) The temperature dependence of the current-voltage characteristics. (b) The raw, non-normalized differential conductance curves at various temperatures below 45 K.

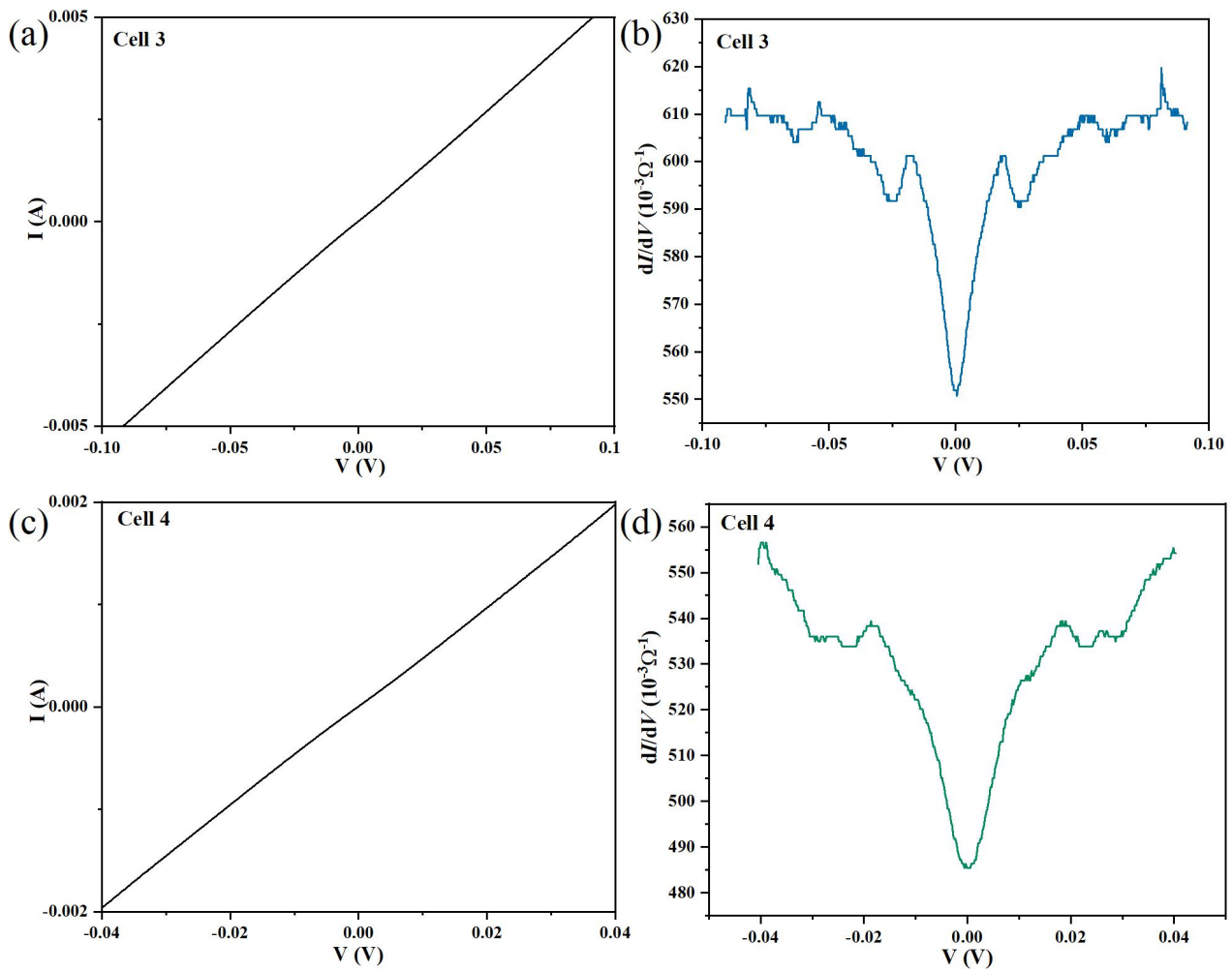

**Extended Data Fig.5. Temperature dependence of the conductance spectra of Cell 3 and Cell4.**
(a) and (c) The temperature dependence of the current-voltage characteristics at 3 K. (b) and (d)The

raw, non-normalized differential conductance curves at 3 K.

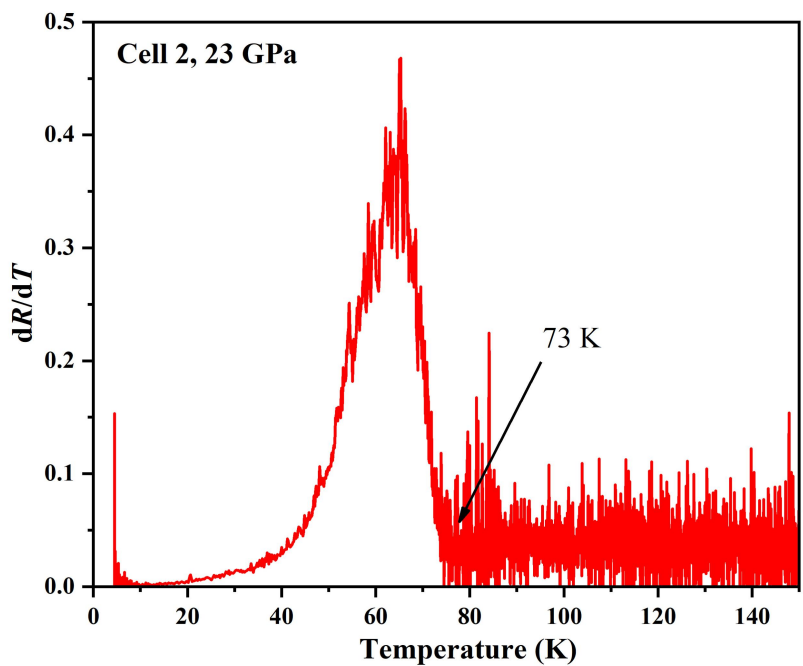

**Extended Data Fig.6. The first derivative of the R-T curve from Cell 2**

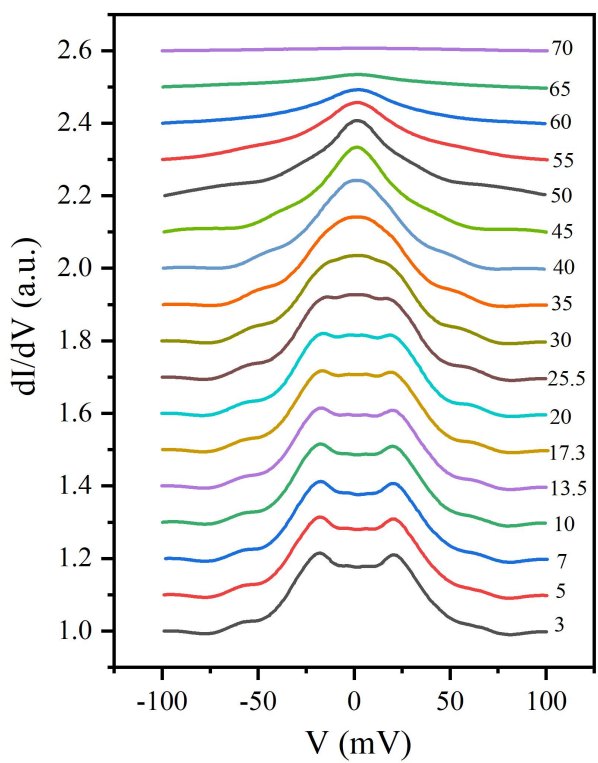

**Extended Data Fig.7. The differential conductance curves of Cell 2, Position 2 in the temperature range of 3 K to 70 K**

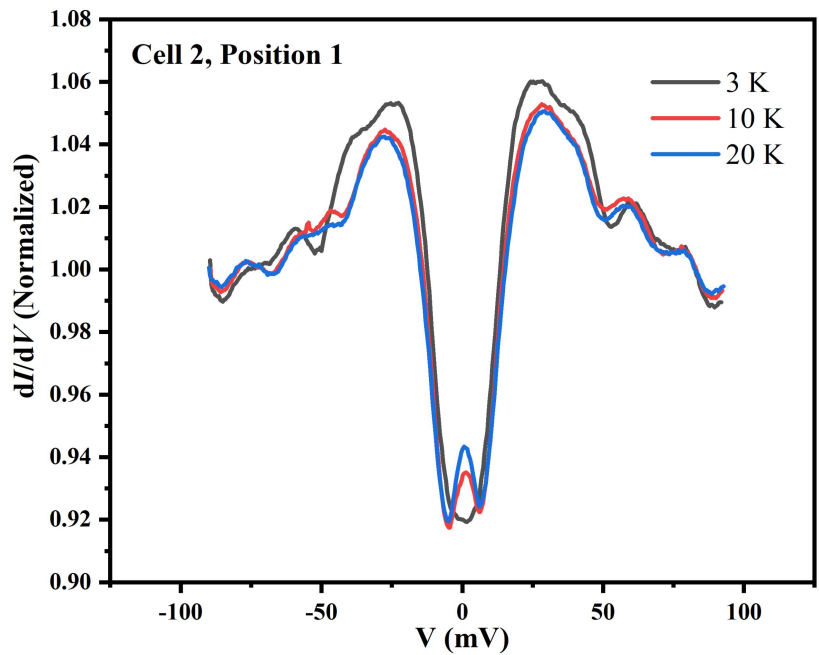

**Extended data Fig.38.** The conductance spectra for Cell 2, position1 along with temperature. Different temperatures are represented by curves in different colors.

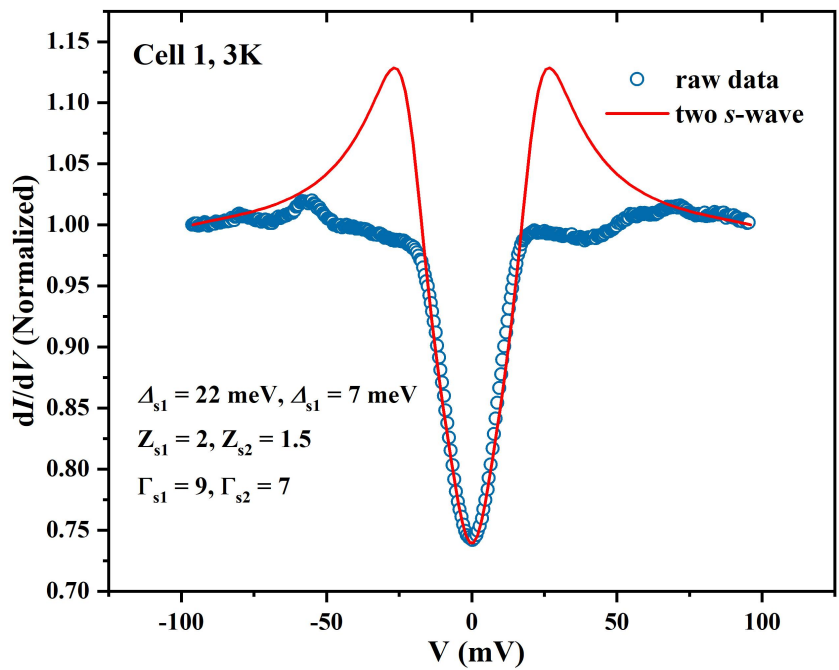

**Extended data Fig.9.** The normalized conductance spectra for Cell 1 at 3 K which matched by the two *s*-wave model.

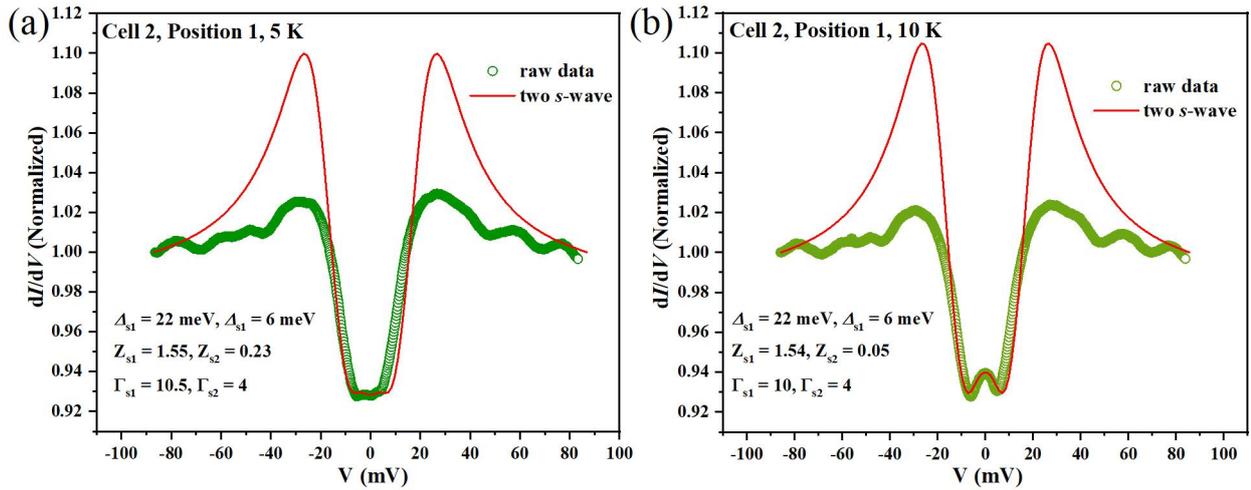

**Extended data Fig.10. The normalized conductance spectra for Cell 2, Position 1 at 5 K and 10 K which matched by the two *s*-wave model**.

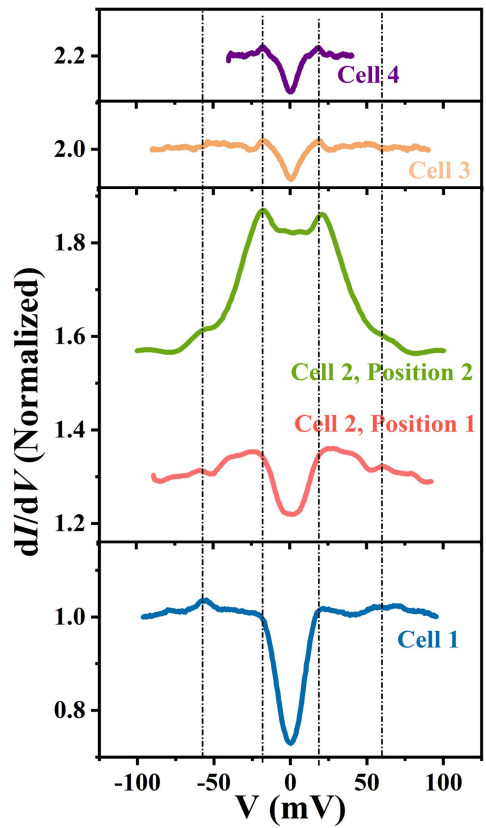

**Extended data Fig.11. Variation of differential conductance with voltage at different positions.** At 3 K, the variation of differential conductance with voltage at different positions of the sample in different cells, represented by different colors. The measurement range is within the sample's critical current.

**Extended Data Table 3. The BTK fit parameters for differential conductance curve of Cell 2, Position 2 at various temperature.**

| Temperature (K) | $\Gamma_1$ (meV) | $\Gamma_2$ (meV) | $Z_1$ | $Z_2$ | $\Delta_1$ (meV) | $\Delta_2$ (meV) |
|---|---|---|---|---|---|---|
| 3 | 6.6 | 4 | 0.58 | 0.4 | 23 | 6 |
| 5 | 6.63 | 4.1 | 0.57 | 0.4 | 22.9 | 6 |
| 10 | 6.63 | 4.1 | 0.557 | 0.41 | 23 | 5.98 |
| 14 | 6.63 | 4.1 | 0.53 | 0.4 | 22.8 | 5.9 |
| 20 | 6.4 | 3.5 | 0.515 | 0.38 | 22.9 | 5.85 |
| 26 | 6.4 | 3.6 | 0.485 | 0.36 | 22.2 | 5.8 |
| 30 | 6.9 | 4 | 0.43 | 0.31 | 22.1 | 5.7 |
| 35 | 7 | 4.1 | 0.4 | 0.3 | 21.5 | 5.6 |
| 40 | 7.2 | 4.3 | 0.35 | 0.27 | 20.5 | 5.3 |

**References**

1. Blonder GE, Tinkham M, Klapwijk TM. Transition from metallic to tunneling regimes in superconducting microconstrictions: Excess current, charge imbalance, and supercurrent conversion. *Physical Review B* **25**, 4515-4532 (1982).
2. Pecchio P, *et al*. Doping and critical-temperature dependence of the energy gaps in Ba(Fe$_{1-x}$Co$_x$)$_2$As$_2$ thin films. *Physical Review B* **88**, 174506 (2013).